\documentclass[AMA,STIX2COL]{MRM}
\articletype{Research Article}%
\input{shortcuts}
\historydates{} 
\doiheadtext{} 
\received{}
\revised{}
\accepted{}
\usepackage{mathtools}
\usepackage{needspace}
\usepackage{amsmath}
\usepackage{adjustbox}        
\usepackage{caption}
\usepackage{subcaption}
\usepackage{enumitem}
\usepackage{pdfcomment}
\usepackage{etoolbox}
\usepackage[capitalise,nameinlink]{cleveref}
\usepackage{marginnote}

\reversemarginpar

\usepackage[dvipsnames]{xcolor}

\makeatletter
\patchcmd{\@maketitle}{\thispagestyle{empty}}{\thispagestyle{titlepage}}{}{}
\patchcmd{\maketitle}{\thispagestyle{titlepage}}{\relax}{}{}
\makeatother

\newcounter{algo}
\renewcommand{\thealgo}{\arabic{algo}}

\algnewcommand{\Constant}{\item[\textbf{Constants:}]}



\newif\ifshowrevs
\showrevsfalse   

\begin{document}

\title{Fully 3D Unrolled Magnetic Resonance Fingerprinting Reconstruction via Staged Pretraining and Implicit Gridding}

\author[1]{Yonatan Urman}{\orcid{0000-0002-5763-8174}}

\author[1]{Mark Nishimura}{\orcid{0000-0003-3976-254X}}

\author[2]{Daniel Abraham}{\orcid{0009-0004-4299-3132}}

\author[1,2]{Xiaozhi Cao}{\orcid{0000-0001-5095-648X}}

\author[1,2]{Kawin Setsompop}{\orcid{0000-0003-0455-7634}}

\authormark{URMAN \textsc{et al}}

\address[1]{\orgdiv{Electrical Engineering}, \orgname{Stanford University}, \orgaddress{\state{California}, \country{USA}}}

\address[2]{\orgdiv{Radiology}, \orgname{Stanford University}, \orgaddress{\state{California}, \country{USA}}}

\corres{Yonatan Urman: \email{yurman@stanford.edu}}
\abstract[Abstract]{
\textbf{Purpose:}
Magnetic Resonance Fingerprinting (MRF) enables rapid quantitative imaging, but high-resolution 3D reconstructions remain computationally expensive due to repeated NUFFTs, and the commonly used Locally Low Rank (LLR) regularization becomes ineffective at high acceleration. Learned 3D priors could address these limitations, but training them at scale is challenging due to memory and runtime constraints. This work proposes SPUR-iG, a fully 3D deep unrolled subspace reconstruction framework that makes large-scale non-Cartesian 3D MRF computationally feasible.

\textbf{Methods:}
SPUR-iG leverages implicit GROG-based data consistency (DC), which grids non-Cartesian k-space using a learned family of kernels, enabling efficient FFT-based DC with minimal artifacts. To enable scalable 3D unrolled training, we introduce a staged training strategy that keeps computation tractable while progressively improving reconstruction quality. We evaluate the method on a large in vivo dataset, as well as on cross-vendor out-of-distribution data.

\textbf{Results:}
At 1 mm isotropic resolution, SPUR-iG outperforms LLR and a state-of-the-art hybrid 2D-3D unrolled baseline in both subspace coefficient quality and $T_1$/$T_2$ accuracy. Whole-brain reconstructions complete in under 15 seconds, providing up to a 111$\times$ speedup for 2-minute scans relative to LLR. Notably, SPUR-iG reconstructions from 30-second acquisitions achieve mean $T_1$ accuracy that match or exceed the mean accuracy of LLR reconstructions from 2-minute acquisitions.

\textbf{Conclusion:}
SPUR-iG introduces a fully 3D unrolled reconstruction framework for MRF that improves both reconstruction speed and accuracy, making high-resolution accelerated 3D MRF more practical for research and clinical use.
\vspace{-4em}

}
\keywords{Deep Learning, Image reconstruction, Magnetic Resonance Fingerprinting, Quantitative MRI, Unrolled networks\vspace{-2em}}

\maketitle

\noindent\ignorespaces

\jnlcitation{\cname{%
\author{Urman Y}, 
\author{Nishimura M}, 
\author{Abraham D}, 
\author{Cao X}, and 
\author{Setsompop K}} (\cyear{2026}), 
\ctitle{Fully 3D Unrolled Magnetic Resonance Fingerprinting Reconstruction via Staged Pretraining and Implicit Gridding}, \cjournal{Magn. Reson. Med.}.}

\section{Introduction}\label{sec:intro}
Quantitative MRI (qMRI) measures intrinsic tissue properties such as $T_1$ and $T_2$, making it less sensitive to acquisition conditions than conventional contrast-weighted imaging, and thereby enabling consistent comparisons across subjects, time points, and sites~\cite{gracien2020stable}. As a result, qMRI has shown promise in studying brain development~\cite{mezer2013quantifying, yablonski2025fast}, and tracking subtle neurodegenerative changes~\cite{tang2018magnetic, luo2019application}, among other applications~\cite{granziera2021quantitative, seiler2021multiparametric}. However, qMRI acquisitions are typically slow since they require sampling the temporal signal evolution of each voxel rather than acquiring a single contrast-weighted snapshot.

Magnetic Resonance Fingerprinting (MRF)~\cite{ma2013magnetic} is a qMRI technique that encodes tissue parameters through a sequence of varying flip angles (FA), repetition times (TR), and potentially other acquisition settings across $\ntr$ time points. At each TR, an image is acquired using an incoherency-promoting trajectory, producing artifacts that are spatially and temporally incoherent and thus reduce interference with tissue-specific signal evolutions. After reconstructing the time series of images, voxel-wise signal evolutions are matched to a dictionary of Bloch-simulated signals, linking tissue parameters to their expected dynamics and enabling parameter estimation.

While highly efficient to acquire, 3D MRF data is computationally demanding to reconstruct, especially at high resolutions~\cite{cao2022optimized, schauman2025deep}. For example, reconstructing an MRF acquisition of the whole
brain with $\ntr=500$ at 1~mm isotropic resolution requires tens of gigabytes to store the time series alone. Moreover, the non-Cartesian reconstruction necessitates repeated non-uniform FFTs (NUFFTs), which can be particularly slow for large 3D problems.

Subspace reconstruction~\cite{zhao2018improved, asslander2018low, liang2007spatiotemporal} approximates the signal dictionary with a low-rank subspace of size $k \ll \ntr$, thereby lowering dimensionality and implicitly regularizing the reconstruction.
The resulting subspace coefficient maps provide a compact representation of the image's signal evolutions across TRs, which can be used to extract quantitative $T_1$ and $T_2$ maps. Beyond traditional parameter mapping, these coefficient maps capture rich signal information that facilitates additional downstream tasks, such as multi-compartment modeling~\cite{nagtegaal2020fast}, and high-quality clinical contrast synthesis~\cite{yurt2024unlocking, wang2023high}. This motivates focusing on accurate subspace coefficient reconstruction that can serve as input to many downstream applications.

For highly accelerated MRF, additional regularization beyond the subspace model is required. Locally Low Rank (LLR)~\cite{tamir2017t2, lima2019sparsity} is widely used, enforcing low-rank structure within local patches, but it requires computationally expensive singular value decompositions (SVD)~\cite{cai2010singular}. At very high acceleration rates, such as acquisitions targeting 1~mm isotropic whole-brain mapping in under 2~minutes, LLR alone is insufficient to preserve reconstruction quality.

Broadly, achieving high-fidelity 3D MRF at these resolutions and acceleration factors is limited by two primary bottlenecks: (1) the reliance on handcrafted priors that lack the expressivity to resolve fine details under high undersampling, and (2) the computationally expensive NUFFTs and regularization operators, which substantially raise reconstruction time and memory usage at high resolutions.

Recent advances address parts of these problems. iGROG~\cite{abraham2023implicit, abraham2024implicit} enables efficient coil-based gridding of non-Cartesian data onto a Cartesian grid using an implicit neural representation (INR)~\cite{mildenhall2021nerf} of the gridding kernel, replacing expensive NUFFTs with efficient FFTs at negligible quality loss. This reduces high-resolution MRF reconstruction times from roughly 30 minutes to around 4 minutes.

To overcome limitations of the LLR prior, state-of-the-art unrolled methods such as SOTIP~\cite{zou2025improved} incorporate learned priors via deep learning, yielding substantial gains in image quality. To accelerate reconstruction, SOTIP employs NUFFTs with reduced kernel sizes and oversampling factors, which introduce small artifacts that the denoiser must learn to remove. Additionally, although data consistency (DC) is performed in 3D, denoising is applied slice-by-slice in 2D to reduce computational cost.

While such design choices are often necessary for computational efficiency, slice-wise denoising limits the ability to fully exploit the inherently 3D structure of MRF data, which exhibits volumetric artifacts and spatial correlations. More generally, several works have explored strategies to enable fully 3D learned reconstruction under memory constraints. Memory-efficient unrolled frameworks based on reversible networks and gradient checkpointing have demonstrated 3D structural MRI reconstruction with substantially reduced memory footprints, at the expense of increased computation and additional numerical considerations in reversible DC and model layers~\cite{kellman2020memory, wang2021memory}. Other approaches reduce memory by reformulating the reconstruction. Efficient Cartesian methods exploit separable structure to process smaller reformatted volumes~\cite{deng2021efficient}, while greedy layer-wise training schemes perform backpropagation after each unroll iteration~\cite{ozturkler2022gleam}, reducing memory by roughly a factor equal to the number of unroll steps, while keeping the overall computational cost approximately unchanged. However, because intermediate iterates are supervised independently rather than through the final unrolled objective, the optimization target may not fully align with the final reconstruction quality. Spatiotemporal architectures have further extended learned reconstruction to Cartesian dynamic imaging by factoring spatial and temporal processing to improve memory utilization~\cite{kustner2020cinenet}.

While these approaches represent important advances, the challenges are further compounded in high-resolution MRF, where the reconstruction problem is inherently 3D, spatiotemporal, and non-Cartesian, limiting the direct application of the above methods. 

To address these issues, we propose SPUR-iG (\textbf{S}taged \textbf{P}retraining for \textbf{U}nrolled \textbf{R}econstruction with \textbf{iG}ROG), a fully 3D deep unrolled reconstruction framework~\cite{aggarwal2018modl, adler2018learned, hammernik2018learning, sriram2020end}. Our approach combines efficient iGROG-based DC with a learned 3D prior and a staged training strategy that enables large-scale unrolled training. We first pretrain a standalone denoiser with extensive data augmentation, then train with a greedy per-unroll-iteration loss to reduce memory demands~\cite{ozturkler2022gleam}, and finally fine-tune the full unrolled model using gradient checkpointing~\cite{chen2016training}. To further improve robustness, the denoiser takes as input both the image and the unroll iteration index, enabling it to amortize performance across varying artifact and noise levels. Evaluated on 45 in vivo MRF acquisitions with retrospective undersampling spanning acquisition times from 2~minutes down to 30~seconds, and including an example from an out-of-distribution scanner vendor, our framework achieves whole-brain 1~mm isotropic reconstructions in under 15~s in all cases, while surpassing existing methods in quality.

The contributions of this work are as follows:
\begin{enumerate}
    \item We present a \emph{fully 3D} unrolled subspace reconstruction method for MRF that improves both the subspace coefficient map quality and quantitative accuracy compared with baseline methods.
    \item We achieve 1~mm whole-brain subspace MRF reconstructions in under 15~seconds by combining efficient iGROG-based DC with a model-based 3D denoiser.
    \item We develop a staged training strategy that enables large-scale 3D unrolled training on moderate hardware and validate it on a large in vivo dataset with diverse acceleration levels, including data from an out-of-distribution scanner vendor.
\end{enumerate}

\section{Methods}
\begin{figure*} \centering
    \includegraphics[width=1\linewidth]{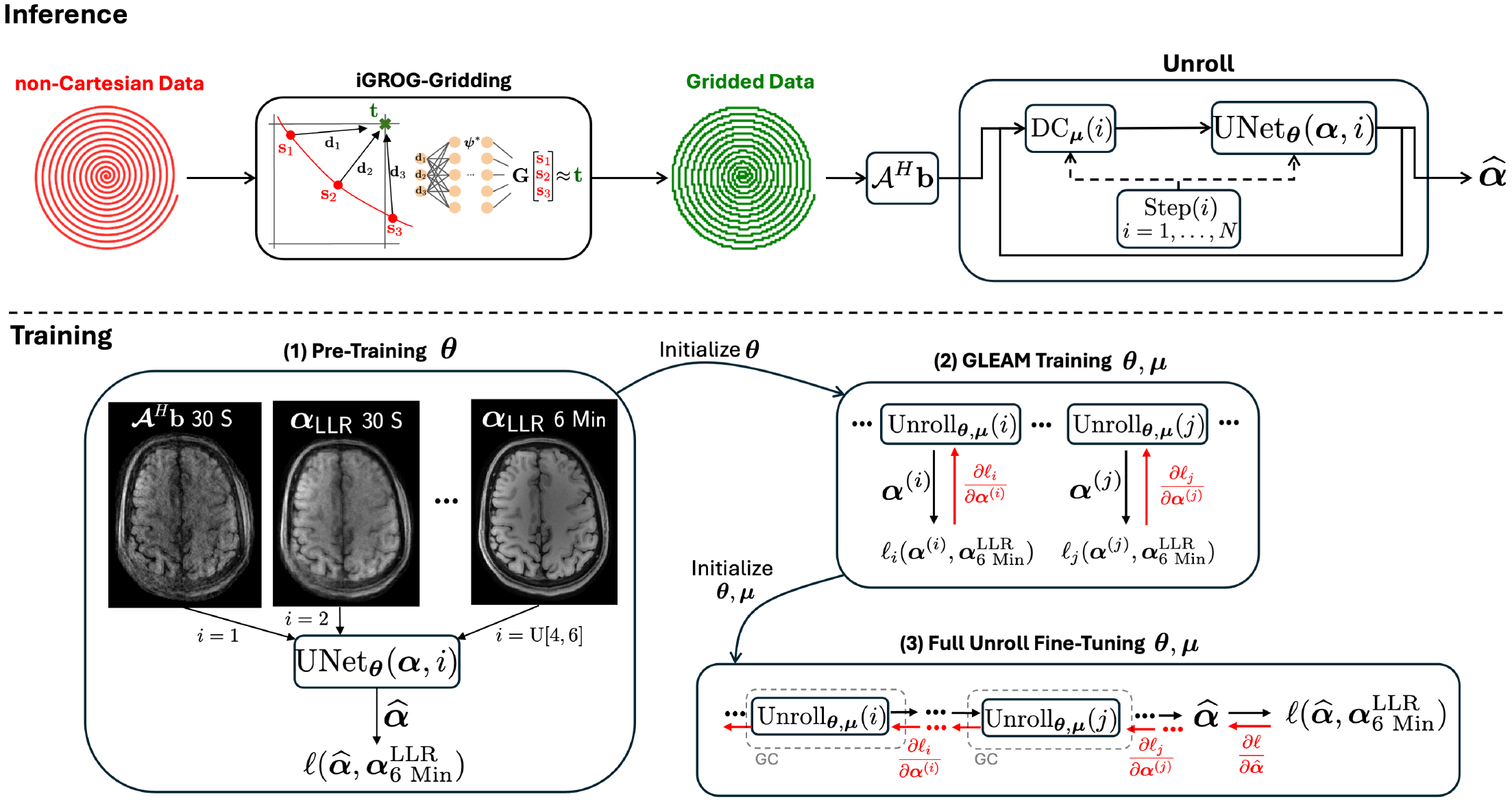}
    \caption{~Illustration of inference (top) and training (bottom) pipelines of SPUR-iG. During inference, iGROG grids the non-Cartesian data onto a Cartesian grid using an implicit kernel representation learned from the calibration region. The gridded k-space is then passed to the unrolled reconstruction, initialized with $\bcA^H \bb$. Reconstruction proceeds for $N$ unroll iterations, each alternating between a DC update with a learnable step size $\mu_i$ and a step-conditioned denoiser. Training is performed in three stages: (1) a step-conditioned UNet is pretrained on inputs with varying artifact levels, using a heuristic mapping between input type and unroll step to promote robustness across noise/artifact conditions; (2) the pretrained UNet is used to initialize a memory-efficient GLEAM-style training, where losses are computed and backpropagated after each unroll iteration, detaching subsequent steps; and (3) full unrolled training, initialized from stage (2), fine-tunes the model under the exact inference setting, with gradient checkpointing (GC) across DC and denoising steps to reduce memory consumption.}
    \label{fig:method}
\end{figure*}

To enable fast and accurate MRF reconstruction, we build on the standard subspace formulation and address its limitations of suboptimal priors and long reconstruction times. We propose a fully 3D unrolled framework that integrates a learned prior with an efficient DC formulation, along with a staged training strategy that makes large-scale training feasible within a reasonable compute and time budget. We begin by reviewing MRF acquisition and the subspace model, followed by unrolled reconstruction, and then present the details of our proposed method and evaluation pipeline. An overview of our approach is illustrated in Figure~\ref{fig:method}.

\subsection{MRF Subspace Reconstruction}
MRF consists of a sequence of $\ntr$ acquisitions with varying excitation parameters (e.g., FAs), producing a time series of images $\bx \in \C^{\ntr \times N_v}$, where $N_v$ denotes the number of voxels in the volume. Reconstruction typically follows two stages: (1) recovery of the time series $\hat{\bx}$ from undersampled k-space data, and (2) voxel-wise parameter estimation via dictionary matching, where the temporal signal evolution is compared to a precomputed dictionary, and quantitative maps are obtained by selecting the entry with maximum cosine similarity.

Directly reconstructing the full time series $\bx$ is computationally intensive and ill-posed due to its high dimensionality and the heavy undersampling used in MRF. This can be mitigated by exploiting temporal redundancy and projecting the data onto a low-dimensional subspace $\bPhi \in \C^{\ntr \times k}$ ($k \ll \ntr$) spanned by the top $k$ singular vectors of the MRF dictionary. The time series is then approximated as $\bx \approx \bPhi \balpha$, where $\balpha \in \C^{k \times N_v}$ are the subspace coefficient maps. We define the forward operator acting on $\balpha$ as $\bcA = \bcF_u \bS \bPhi$, with $\bS$ the coil sensitivities and $\bcF_u$ the undersampled Fourier operator. Since $\bcF_u \bS$ acts on the spatial domain and $\bPhi$ on the temporal domain, these operators commute, allowing the costly $\bcF_u \bS$ to be applied directly in the reduced $k$-dimensional space~\cite{tamir2017t2}.

At high acceleration, however, the subspace model alone is insufficient, and the reconstruction problem is typically augmented with an LLR prior, leading to
\begin{equation}\label{eq:mrf_llr} 
\hat{\balpha} = \argmin{\balpha}{\|\bcA(\balpha) - \bb\|_2^2 + \lambda \,\text{LLR}(\balpha)},
\end{equation}
where $\bb$ are k-space measurements and $\text{LLR}(\balpha)$ enforces low-rank structure within local spatial patches of the coefficient maps.

\subsection{Unrolled Reconstruction} 
Equation \eqref{eq:mrf_llr} is commonly solved using accelerated proximal algorithms such as FISTA~\cite{beck2009fast}, which alternate between DC updates and proximal steps. However, at very high acceleration rates, LLR regularization becomes insufficient to constrain the undersampled problem, and in addition, its proximal operator requires repeated SVDs of local patches, making it computationally expensive.

Unrolled reconstruction~\cite{aggarwal2018modl, adler2018learned, hammernik2018learning, sriram2020end} addresses these issues by replacing the proximal operator with a learned denoiser, yielding the update scheme
\begin{align} \balpha_{\tfrac{1}{2}}^{(i+1)} &= \balpha^{(i)} - \mu_i \, \bcA^H\!\left(\bcA \balpha^{(i)} - \bb\right), \text{(DC
step)} \label{eq:unroll_dc} \\ \balpha^{(i+1)} &= f_\btheta\!\left(\balpha_{\tfrac{1}{2}}^{(i+1)}\right), \quad\quad\quad\quad\text{(learned
prior)} \label{eq:unroll_prior}
\end{align}
where $f_\btheta$ is a deep learning–based denoiser and $\mu_i$ are learnable step sizes. By unrolling this computation for a fixed number of steps, $f_\btheta$ can be trained to recover clean images from undersampled inputs, enabling it to learn rich image statistics that surpass hand-crafted priors.

\subsection{SPUR-iG Reconstruction}
In large-scale 3D non-Cartesian settings, unrolled methods face two key challenges: (1) DC updates remain slow, as evaluating $\bcA^H \bcA(\cdot)$ in \eqref{eq:unroll_dc} requires repeated NUFFTs, and (2) training fully 3D unrolled networks is memory-intensive, since the entire unrolled computation graph must be stored for backpropagation. We address these challenges with an efficient DC implementation and a scalable, progressive training framework.

\subsubsection{Efficient Data Consistency via iGROG}
The main cost in \eqref{eq:unroll_dc} arises from evaluating $\bcA^{H}\bcA(\cdot)$, which relies on NUFFTs. To accelerate this step, we use iGROG, which learns an INR of the gridding kernel across multi-channel coils. Using calibration data, the INR is trained to map $n$ offset vectors $\bd_1,\ldots,\bd_n$, corresponding to displacements between non-Cartesian samples and a target Cartesian grid point, to a kernel $G_{\bpsi^*}(\bd_1,\ldots,\bd_n)$. For each target grid location, nearby source samples are selected, the kernel is evaluated at their relative displacements, and the resulting kernel weights are used to compute the target k-space value. This enables GRAPPA-like~\cite{griswold2002generalized} interpolation and, through the use of an INR, generalizes across the diverse orientations encountered in the trajectory. This allows us to first grid the non-Cartesian data onto a Cartesian grid, after which NUFFTs can be replaced by efficient FFTs. The upper left part of Figure~\ref{fig:method} illustrates this process.

The approach provides flexibility to trade off accuracy and noise amplification against speed. By using an oversampled grid, interpolation quality improves (since the interpolation distances are smaller), but reconstruction time increases. In our experiments, we used a $\times1.5$ oversampling ratio with 5 k-space data points per coil channel, achieving high-quality reconstructions while retaining substantial speed-ups. Across all acquisitions considered in this work, INR training completed in under 25~s and can be performed during MRF acquisition immediately after calibration data are acquired. Once trained, gridding of the dataset is likewise efficient, completing in under 3~s and yielding data on a Cartesian grid for subsequent FFT-based DC updates.

\subsubsection{Efficient Training of a 3D Unrolled Network}
For $f_\btheta$ in \eqref{eq:unroll_prior}, we use a fully 3D UNet~\cite{ronneberger2015u}, which enables the model to capture volumetric context and learn an effective unrolled denoising strategy. To improve data efficiency, weights are shared across unroll iterations, and FiLM-based~\cite{perez2018film} conditioning is introduced on the iteration index, i.e., $f_\btheta(\balpha, i)$. This conditioning enables the network to adapt its behavior across unroll steps, improving robustness by amortizing training over different noise and artifact regimes, similar to time conditioning in diffusion models~\cite{ho2020denoising, song2020score}.

To enable training of a 3D unrolled network in the challenging MRF reconstruction setting, we adopt a progressive three-stage strategy, illustrated in the lower part of Figure~\ref{fig:method}. This strategy consists of the following stages:
\begin{enumerate}
  \item \textbf{Denoiser pretraining:} We begin by pretraining a standalone, iteration-conditioned 3D UNet on a diverse set of inputs, where the model is trained to recover clean images from noisy, undersampled inputs. To obtain a robust initialization for subsequent stages, we expose the network to a wide range of artifact and noise levels at its input, including zero-filled ($\bcA^H \bb$) and LLR-based reconstructions from scans with different acquisition durations (30~s, 1~min, 2~min, and 6~min, denoted $\balpha_{\text{LLR}}^{\text{30s}}, \ldots, \balpha_{\text{LLR}}^{\text{6min}}$). Since the UNet is iteration-conditioned, but this stage does not involve actual unrolling and therefore has no natural iteration index, we assign heuristic pseudo-iteration indices to each input type. Specifically, $\bcA^H \bb$ is paired with index~1, corresponding to the initialization of unrolled reconstruction, while reconstructions with increasing acquisition durations are mapped to higher indices (e.g., $\balpha_{\text{LLR}}^{\text{30s}} \to 2$). Because this stage does not involve expensive unrolling or DC updates, we can efficiently apply extensive data augmentation, including random rotations, translations, and scaling. At the end of this stage, the network provides a well-initialized denoising prior that generalizes across a range of artifact levels and incorporates a notion of the unroll iteration index.

  \item \textbf{Greedy unrolled training (GLEAM}~\cite{ozturkler2022gleam}\textbf{):} Starting from the pretrained weights, we train the unrolled network using a greedy scheme, in which the loss is backpropagated through only a single unroll iteration at a time, after which the computation graph is detached. This avoids storing the full unrolled graph, which would otherwise exceed typical GPU memory. Since the \emph{final} output of the unrolled process is of primary interest, we assign geometrically increasing loss weights across iterations, emphasizing later steps while allowing greater flexibility in earlier ones. Importantly, this stage resolves the input distribution mismatch introduced during prior pretraining: while the denoiser from Stage~1 is trained on reconstructed images with different artifacts levels, during unrolled inference its true inputs are the intermediate outputs of preceding iterations. By the end of this stage, we obtain a fully unrolled model that achieves high reconstruction quality, with faster training convergence and improved performance due to the informed initialization.
  
  \item \textbf{Final fine-tuning:} We finally fine-tune the model using full unrolled training, initialized from the GLEAM weights. In contrast to GLEAM, where losses applied at intermediate iterations may not fully align with optimal final reconstruction quality, this stage directly optimizes the denoiser with respect to the final output, thereby correcting this mismatch. Since full unrolled training is memory-intensive, we employ gradient checkpointing across both DC and denoising steps to reduce memory usage. Although each update in this stage is slow, the GLEAM-based initialization enables convergence in relatively few training iterations.
\end{enumerate}

Collectively, these stages enable efficient large-scale training by progressively mitigating the limitations of each preceding stage, while circumventing the need to train a fully 3D unrolled model from scratch, which would otherwise be prohibitively compute expensive and slow. This process ultimately yields a high-quality, fully 3D unrolled reconstruction model. We refer to the resulting framework as \textbf{SPUR-iG}, introduced in Section~\ref{sec:intro}. Code will be made publicly available upon publication.

\subsection{Dataset}
We used a FISP-MRF sequence with a 3D Spiral Projection Imaging (SPI) acquisition~\cite{cao2019fast}, beginning with an inversion pulse ($\text{TI}=20$~ms) followed by $\ntr=500$ TRs ($\text{TE}=1.7$~ms, $\text{TR}=12.5$~ms). Each block of 500 TRs was extended by 40 additional TRs to acquire a low-resolution (4~mm isotropic) navigator for motion estimation\cite{cao2025three}. The combined 540 TRs form an acquisition group with a total duration of 7.9~s, including a 1.2~s resting delay between groups to enable $M_z$ recovery. Multiple groups are acquired using different spiral projection interleaves and rotations to build up sufficient k-space encoding per TR position, with sampling patterns optimized for k-space coverage and incoherent aliasing using Tiny Golden-Angle Shuffling (TGAS)~\cite{cao2022optimized}. The acquisition resolution was 1~mm isotropic with a $220~\text{mm} \times 220~\text{mm} \times 220~\text{mm}$ field of view. Two tailored dummy-scan groups (16~s) were acquired at the start for $B_1$ estimation~\cite{cao2025invivo}, and to enable steady-state signal to be reached prior to MRF acquisition.

Following the convention from TGAS~\cite{cao2022optimized}, we define an $R=1$ acquisition as 50 groups (including the two dummy groups), corresponding to a total scan time of 6~min 38~s. We considered undersampling factors of $R=3,6,$ and $12$, yielding scan times of 2~min 23~s, 1~min 19~s, and 47~s, respectively. For reporting, we use the raw MRF acquisition time (excluding the fixed dummy-group overhead), which we round and denote as 6~min, 2~min, 1~min, and 30~s for convenience. Practical strategies to bypass the need for these dummy groups are discussed in Section~\ref{sec:dis_quant}. All scans were coil-compressed to 10 virtual coils, and coil sensitivity maps were estimated using ESPIRiT~\cite{uecker2014espirit}.

The dataset comprised 44 in vivo subjects scanned on a 3T GE SIGNA Premier system, with 36 used for training, 4 for validation, and 4 for testing. To assess cross-vendor generalization, we additionally evaluated our method on a subject scanned on a 3T Siemens MAGNETOM Vida system. For all subjects, motion was corrected in k-space using navigator data~\cite{cao2025three}. The $R=1$ acquisition was reconstructed using subspace reconstruction with light LLR regularization via FISTA and served as the training target. Quantitative maps were subsequently estimated using a $B_1$-corrected dictionary to mitigate systematic $T_2$ bias~\cite{sled2000correction}.

To improve numerical stability in SPUR-iG, which performs joint denoising across subspace coefficient maps, we do not directly use the raw SVD basis $\bPhi$ derived from the MRF signal dictionary, as later components contain substantially less energy. Instead, we apply basis balancing~\cite{schauman2025deep} by performing a discrete Fourier transform along the subspace dimension, $\tilde{\bPhi} = \bF_k(\bPhi)$, where $\bF_k$ denotes a unitary DFT of length $k$. Since $\bF_k$ is orthonormal, this transformation is invertible and leaves the reconstruction solution unchanged, while redistributing energy more evenly across coefficients and thereby stabilizing the training of the learned denoisers. For a fair comparison, the balanced basis was used for all baseline methods.

Finally, we synthesized MPRAGE images from the $R=1$ quantitative maps using Bloch simulation and used them to obtain segmentation masks with FreeSurfer~\cite{fischl2012freesurfer}. These masks were then used to apply region-specific loss weighting during model training (additional details in Appendix~\ref{sec:apx_loss}).


\subsection{Evaluation} 
We evaluated our approach on the held-out test set using retrospectively undersampled acquisitions of 2~min, 1~min, and 30~s. Evaluation considered both the reconstruction quality of the subspace coefficient maps and the resulting quantitative $T_1$/$T_2$ maps, obtained by voxel-wise dictionary matching with $B_1$-corrected dictionaries. In addition, we performed ablation studies to analyze the contribution of different components of our framework.

We benchmarked against two baselines:
\begin{enumerate}
    \item \textbf{LLR}: Subspace reconstruction with FISTA and LLR regularization. The regularization weight was tuned on the validation set for each scan duration to minimize the sum of the mean relative $T_1$ and $T_2$ errors within WM and GM. The number of iterations was fixed at $40$, ensuring convergence.
    \item \textbf{Hybrid 2D/3D}: Inspired by the state-of-the-art SOTIP~\cite{zou2025improved}, this baseline alternates 3D DC updates with axial slice-wise 2D denoising. For a fair comparison, we train this model using the same progressive procedure. Unlike SPUR-iG, the denoiser here is a 2D UNet applied independently to each slice at every unroll iteration, with the 3D volume subsequently formed by stacking the denoised slices. To control for model capacity, we increased the number of parameters in the 2D UNet such that both this model and SPUR-iG saturated the memory of a single NVIDIA A6000 GPU (48~GB) during the final fine-tuning stage
\end{enumerate}
Separate models were trained for each scan duration. For the unrolled models, we used six unroll iterations, selected based on ablation studies (Section~\ref{sec:met_ablation}). Additional implementation details are provided in Appendix~\ref{sec:impl_details}.




\section{Results}

\begin{table*}[htbp]
\centering
\resizebox{\textwidth}{!}{%
\begin{tabular}{clccccc|ccccc}
\toprule
 &  & \multicolumn{5}{c}{PSNR (dB) $\uparrow$} & \multicolumn{5}{c}{SSIM $\uparrow$} \\
 & Method & $c_{1}$ & $c_{2}$ & $c_{3}$ & $c_{4}$ & $c_{5}$ & $c_{1}$ & $c_{2}$ & $c_{3}$ & $c_{4}$ & $c_{5}$ \\
\midrule
\multirow{3}{*}{\rotatebox[origin=c]{90}{2 Min}} & SPUR-iG (Ours) & \textbf{33.6$\pm$2.1} & \textbf{32.7$\pm$2.7} & \textbf{32.1$\pm$2.3} & \textbf{32.1$\pm$2.2} & \textbf{33.0$\pm$2.3} & \textbf{0.94$\pm$0.02} & \textbf{0.87$\pm$0.03} & \textbf{0.87$\pm$0.03} & \textbf{0.87$\pm$0.03} & \textbf{0.87$\pm$0.03} \\
 & Hybrid 2D/3D & 32.9$\pm$2.1 & 32.3$\pm$2.5 & 31.4$\pm$2.2 & 31.3$\pm$2.3 & 32.0$\pm$2.4 & 0.93$\pm$0.03 & 0.86$\pm$0.03 & 0.85$\pm$0.03 & 0.86$\pm$0.03 & 0.87$\pm$0.03 \\
 & LLR & 32.0$\pm$1.9 & 31.1$\pm$2.2 & 31.0$\pm$1.9 & 30.8$\pm$1.9 & 30.7$\pm$2.0 & 0.91$\pm$0.03 & 0.83$\pm$0.04 & 0.84$\pm$0.03 & 0.84$\pm$0.03 & 0.83$\pm$0.03 \\
\midrule
\multirow{3}{*}{\rotatebox[origin=c]{90}{1 Min}} & SPUR-iG (Ours) & \textbf{30.4$\pm$2.3} & \textbf{30.5$\pm$2.8} & \textbf{29.7$\pm$2.4} & \textbf{29.2$\pm$2.7} & \textbf{30.4$\pm$2.4} & \textbf{0.89$\pm$0.04} & \textbf{0.79$\pm$0.04} & \textbf{0.80$\pm$0.04} & \textbf{0.80$\pm$0.05} & \textbf{0.80$\pm$0.04} \\
 & Hybrid 2D/3D & 29.5$\pm$2.3 & 29.6$\pm$2.5 & 28.9$\pm$2.3 & 28.7$\pm$2.5 & 29.0$\pm$2.7 & 0.87$\pm$0.04 & 0.78$\pm$0.04 & 0.78$\pm$0.04 & 0.78$\pm$0.04 & 0.78$\pm$0.04 \\
 & LLR & 28.2$\pm$2.1 & 27.9$\pm$2.2 & 28.0$\pm$2.1 & 27.8$\pm$2.1 & 27.5$\pm$1.9 & 0.81$\pm$0.06 & 0.69$\pm$0.05 & 0.73$\pm$0.04 & 0.72$\pm$0.04 & 0.70$\pm$0.04 \\
\midrule
\multirow{3}{*}{\rotatebox[origin=c]{90}{30 S}} & SPUR-iG (Ours) & \textbf{28.0$\pm$2.4} & \textbf{28.5$\pm$3.0} & \textbf{27.6$\pm$2.6} & \textbf{26.9$\pm$2.9} & \textbf{28.8$\pm$2.5} & \textbf{0.84$\pm$0.06} & \textbf{0.75$\pm$0.06} & \textbf{0.75$\pm$0.05} & \textbf{0.74$\pm$0.05} & \textbf{0.75$\pm$0.06} \\
 & Hybrid 2D/3D & 27.2$\pm$2.6 & 27.9$\pm$2.9 & 26.6$\pm$2.7 & 26.0$\pm$3.0 & 27.3$\pm$2.6 & 0.82$\pm$0.06 & 0.72$\pm$0.06 & 0.72$\pm$0.05 & 0.71$\pm$0.05 & 0.73$\pm$0.06 \\
 & LLR & 24.2$\pm$2.2 & 24.7$\pm$2.4 & 24.0$\pm$2.2 & 23.9$\pm$2.3 & 24.5$\pm$2.0 & 0.72$\pm$0.07 & 0.57$\pm$0.06 & 0.61$\pm$0.05 & 0.60$\pm$0.05 & 0.56$\pm$0.05 \\
\bottomrule
\end{tabular}%
}
\caption{~PSNR (left) and SSIM (right) across the five balanced subspace coefficients (c1–c5). Results are shown for three scan durations: 2~min (top), 1~min (middle), and 30~s (bottom). Metrics are computed on axial brain slices (excluding non-brain slices) and averaged across slices from all test subjects. Values are reported as mean $\pm$ standard deviation across slices. Bold entries indicate values that are statistically significantly larger than those of the other methods.}
\label{tab:combined_subspace}
\end{table*}

\begin{figure*} \centering
    \begin{subfigure}[t]{0.5\linewidth} \centering
        \includegraphics[width=\linewidth]{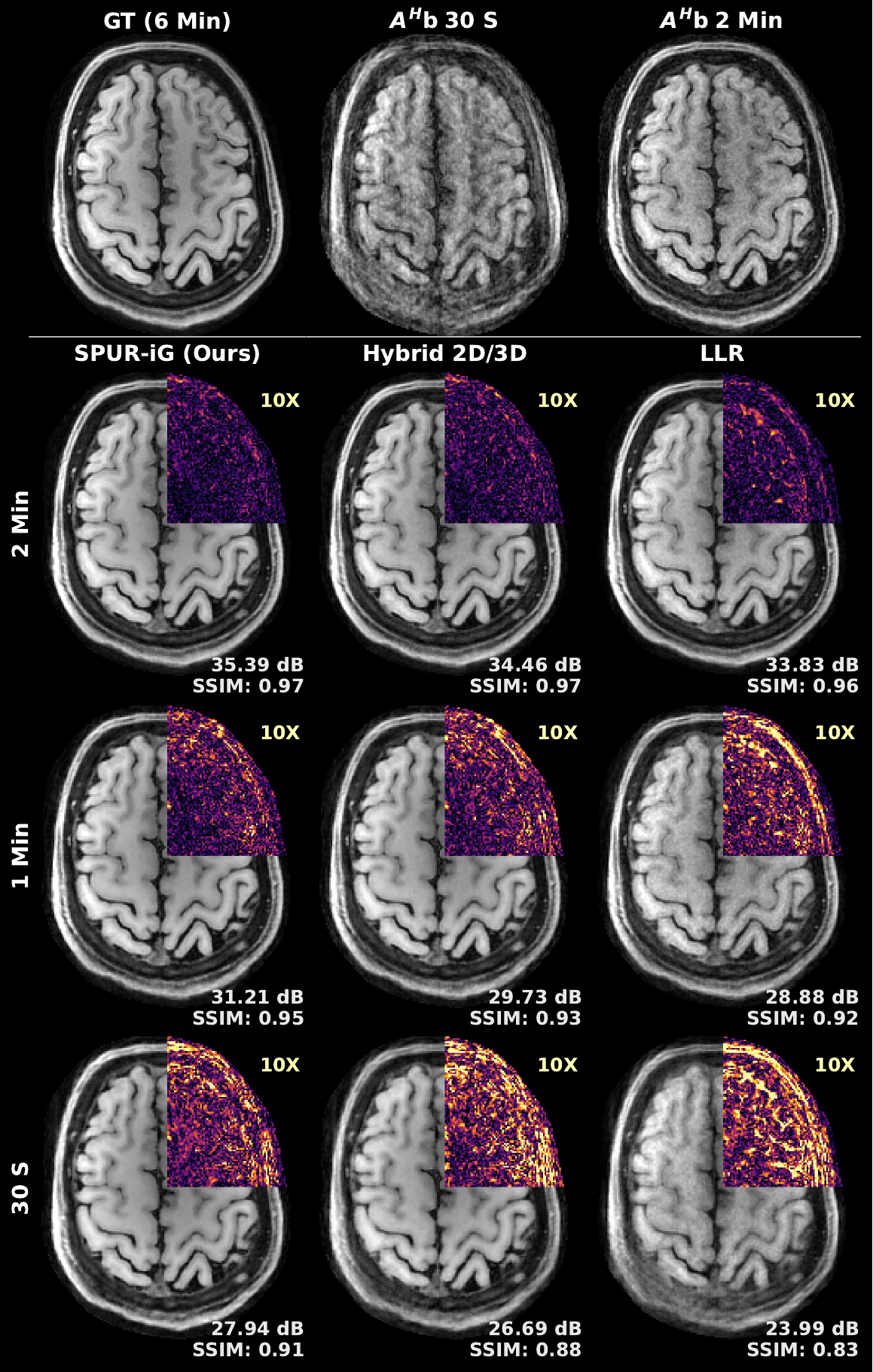}
        \caption{}\label{fig:subspace_ex}
    \end{subfigure}
    \begin{subfigure}[t]{0.493\linewidth} \centering
        \includegraphics[width=\linewidth]{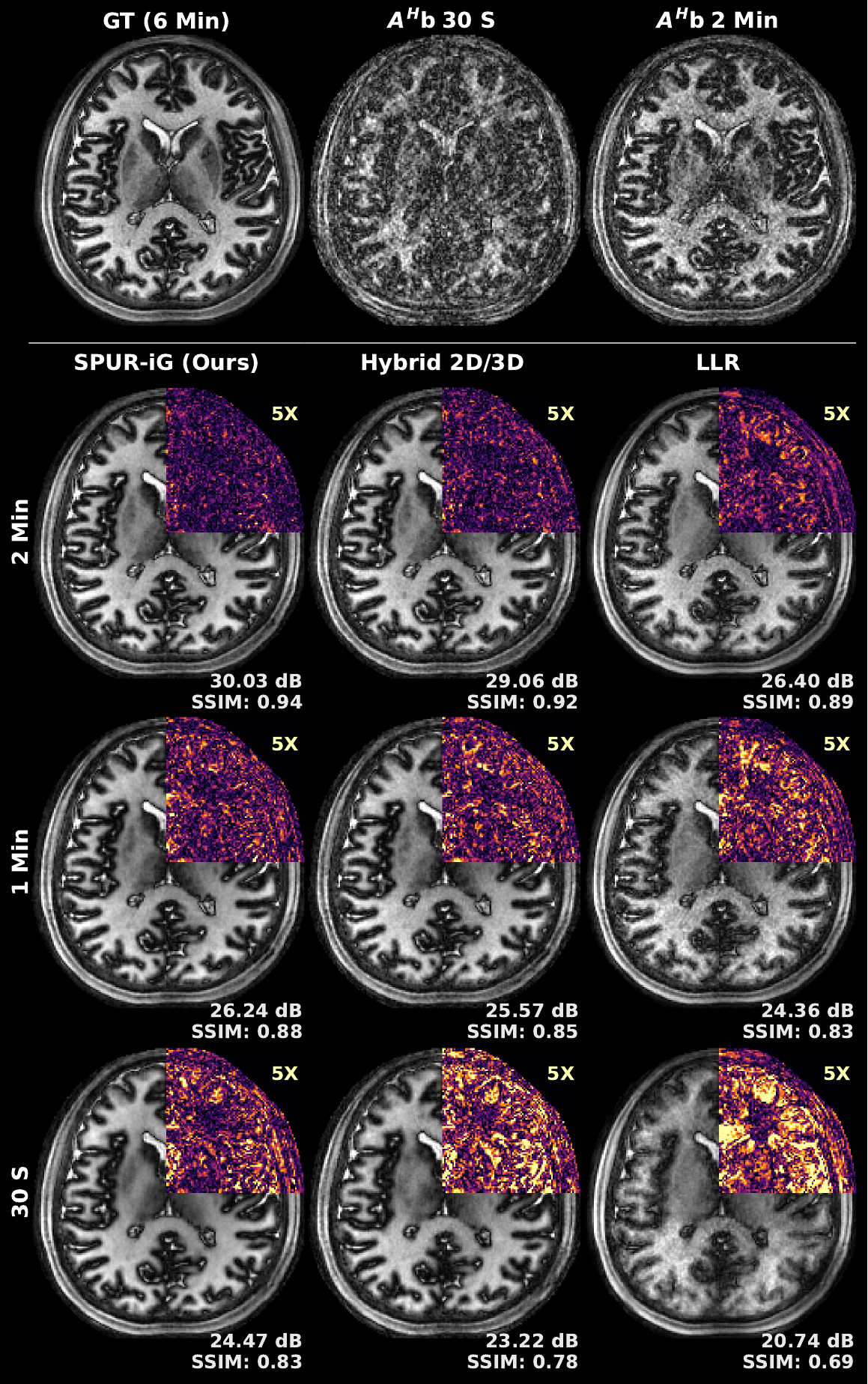}
        \caption{}\label{fig:unbalanced_subspace_ex}
    \end{subfigure}
    \caption{~Example reconstructions of the first balanced subspace coefficient (a) and third unbalanced coefficient (b) for two test subjects. The top row shows the 6~min reference (left) and $\bcA^H\bb$ for the 30~s and 2~min acquisitions (middle and right) as initialization references. Subsequent rows correspond to reconstructions at 2~min, 1~min, and 30~s, with columns comparing our method, the hybrid 2D/3D variant and LLR. PSNR and SSIM are reported in the bottom right of each image. Error maps of the top-right quadrant (magnified $\times 10$ or $\times 5$) are overlaid to illustrate error levels.}\label{fig:subspace_combined}
\end{figure*}

\subsection{Coefficient Reconstruction} 
Table~\ref{tab:combined_subspace} shows the PSNR and SSIM for the five balanced subspace coefficients across scan durations and reconstruction methods. SPUR-iG consistently achieves higher PSNR and SSIM compared to both LLR and the hybrid 2D/3D method, with performance gaps widening as scan duration decreases. Statistical testing confirms that our method outperforms both baselines with $p<0.001$ across all metrics and subspace dimensions.

Representative qualitative results are shown in Figure~\ref{fig:subspace_combined} for the first balanced and third unbalanced subspace coefficients. These were chosen to display since balanced coefficients generally display flatter contrast, whereas unbalanced coefficients, particularly the third, exhibit higher tissue contrast and reveal finer anatomical details. As scan time decreases, all methods exhibit increased noise, but our approach shows the smallest degradation. For the 30~s scans, both LLR and the hybrid 2D/3D variant produce noisy and over-smoothed reconstructions, whereas our method better preserves fine structure and anatomical details.

\subsection{Quantitative Reconstruction} 
Table~\ref{tab:quant_accuracy} shows the average relative quantitative $T_1$ and $T_2$ errors inside the brain (excluding CSF), defined as $\lvert T^{\text{GT}} - \widehat{T} \rvert / T^{\text{GT}}$ for $T \in \{T_1, T_2\}$, where $\widehat{T}$ denotes the estimated parameter. Across all scan durations, our method achieves lower errors than both baselines. As expected, accuracy decreases with shorter acquisitions, but degradation is less pronounced for our method. Notably, for $T_1$, our reconstructions of the 30~s scans achieved lower average error than LLR reconstructions of 2~min scans. For $T_2$, our approach also provides consistent improvements. Statistical testing confirms that our method outperforms both baselines with $p<0.001$ for $T_1$ and $T_2$ across all scan durations.

\begin{figure*} \centering
    \begin{subfigure}[t]{0.5\linewidth} \centering
        \includegraphics[width=\linewidth]{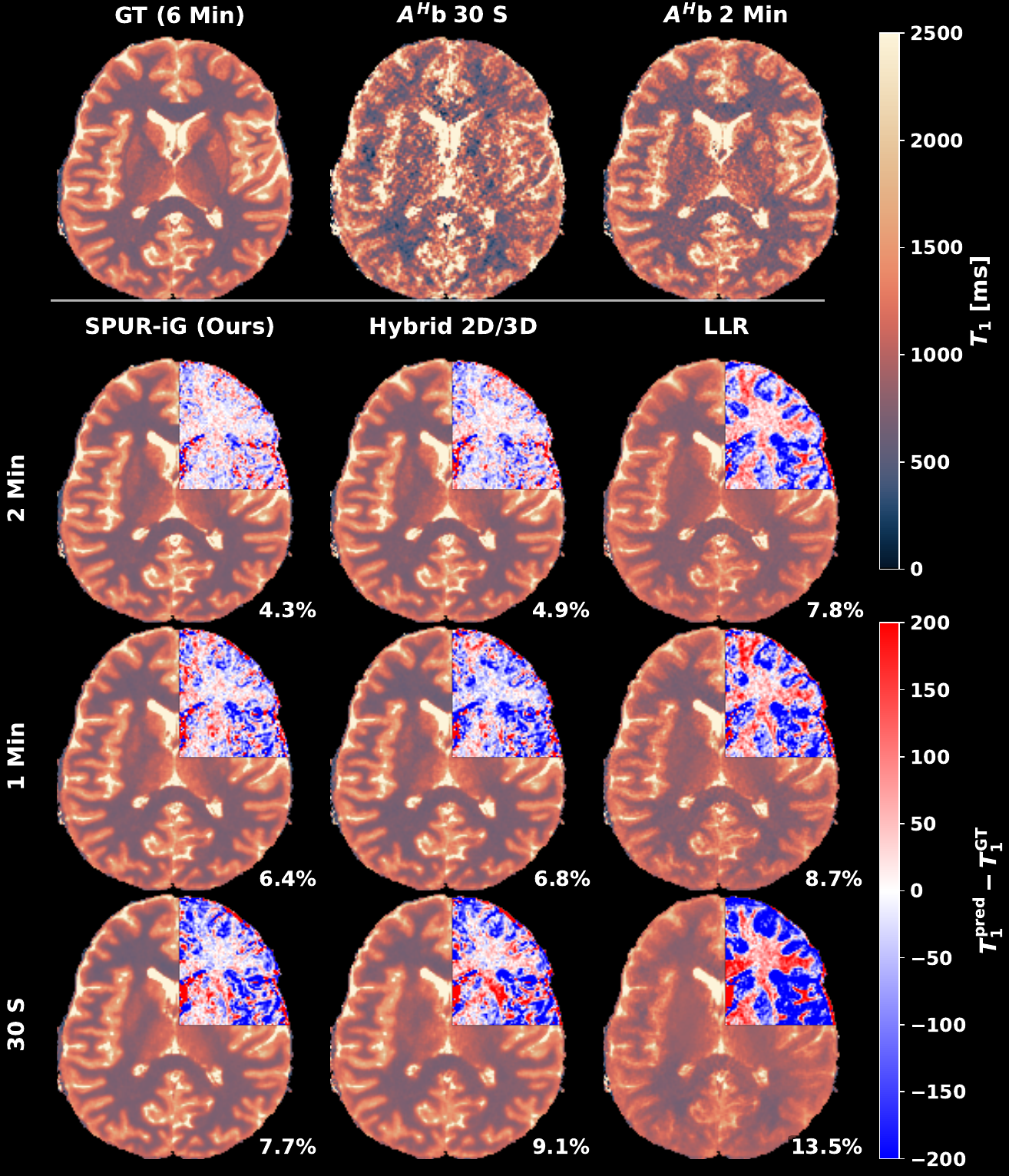}
        \caption{$T_1$ maps}
        \label{fig:quant_t1_ex}
    \end{subfigure}
    \begin{subfigure}[t]{0.493\linewidth} \centering
        \includegraphics[width=\linewidth]{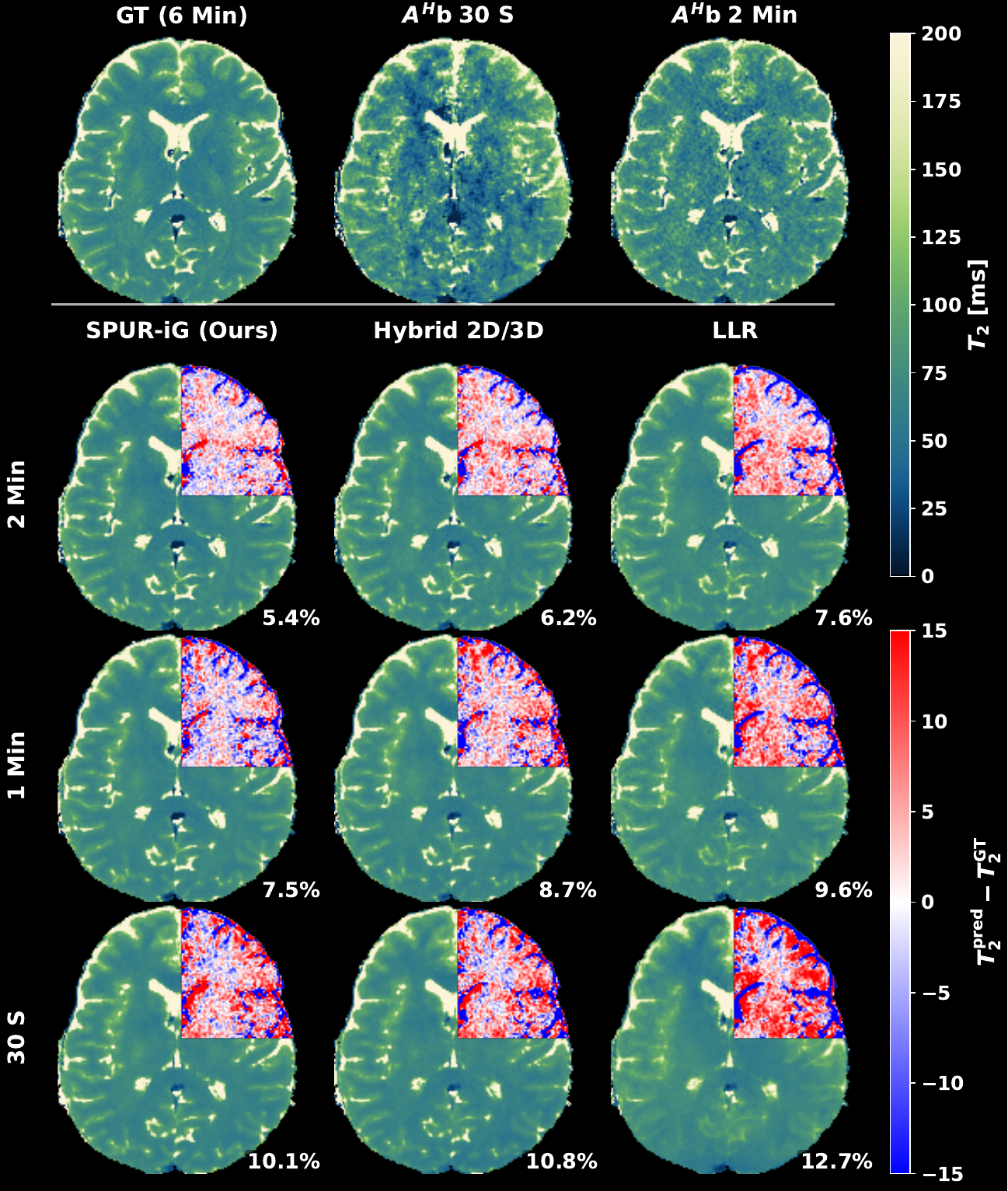}
        \caption{$T_2$ maps}
        \label{fig:quant_t2_ex}
    \end{subfigure}
    \caption{~Qualitative examples of reconstructed $T_1$ (a) and $T_2$ (b) maps. The first row shows the 6~min reference (left) and maps from $\bcA^H \bb$ for the 30~s and 2~min acquisitions (middle and right) as references. Subsequent rows present reconstructions at 2~min, 1~min, and 30~s, with columns comparing our method, the hybrid 2D/3D variant, and LLR. Error maps of the top-right quadrant are overlaid on each image. Quantitative colormap (top) and error colormap (bottom) are shown to the right.}
    \label{fig:quant_t1t2_ex}
\end{figure*}

Representative qualitative results are presented in Figure~\ref{fig:quant_t1t2_ex}. For 2~min scans, all methods produce reasonable maps, with ours showing the lowest residual errors. At 30~s, the baseline methods yield overly smoothed maps that lose fine anatomical details, whereas SPUR-iG preserves sharper structures and reduces error.


\begin{table*}[htbp]
\centering
\begin{tabular}{lcc|cc|cc}
\toprule
 & \multicolumn{2}{c}{2 Min} & \multicolumn{2}{c}{1 Min} & \multicolumn{2}{c}{30 S} \\
Method & $T_1$ $\downarrow$ & $T_2$ $\downarrow$ & $T_1$ $\downarrow$ & $T_2$ $\downarrow$ & $T_1$ $\downarrow$ & $T_2$ $\downarrow$ \\
\midrule
SPUR-iG (Ours) & \textbf{3.9$\pm$2.9} & \textbf{4.8$\pm$3.6} & \textbf{5.6$\pm$4.1} & \textbf{6.7$\pm$4.9} & \textbf{7.1$\pm$5.2} & \textbf{8.9$\pm$6.5} \\
Hybrid 2D/3D & 4.4$\pm$3.2 & 5.1$\pm$3.8 & 6.2$\pm$4.5 & 7.5$\pm$5.5 & 7.7$\pm$5.6 & 9.5$\pm$6.9 \\
LLR & 7.2$\pm$5.0 & 6.4$\pm$4.5 & 8.2$\pm$5.6 & 8.3$\pm$5.9 & 11.3$\pm$7.3 & 11.1$\pm$8.2 \\
\bottomrule
\end{tabular}
\caption{~Relative quantitative $T_1$ and $T_2$ errors across methods and scan durations. Values are reported as mean $\pm$ standard deviation across voxels, averaged over all test subjects, and computed within the brain (excluding CSF). Bold entries indicate values that are statistically significantly lower than those of the other methods.}
\label{tab:quant_accuracy}
\end{table*}

\subsection{Method Ablation}\label{sec:met_ablation}
To assess the contribution of individual training stages, we compare the following variants:
\begin{enumerate}
    \item \textbf{Full}: Our complete approach with all three training stages.
    \item \textbf{GLEAM}: The model after the first two stages, before final fine-tuning.
    \item \textbf{GLEAM (no PT)}: A model trained with GLEAM from scratch, without
    the pretraining stage. Training time is matched to the combined pretraining and GLEAM
    training of the \textbf{GLEAM} variant for fair comparison.
    \item \textbf{GLEAM x3 (no PT)}: Same as \textbf{GLEAM (no PT)}, but trained for three times longer.
\end{enumerate}

To evaluate the effect of unroll iteration conditioning and weight sharing, we further compare:
\begin{enumerate}
    \setcounter{enumi}{4}
    \item \textbf{Full (uncond.)}: Full training using an unconditioned model, i.e., without unroll iteration conditioning.
    \item \textbf{Full (uncond., no WS)}: Same as \textbf{Full (uncond.)}, but without weight sharing in the GLEAM and fine-tuning stages (resulting in a $6\times$ increase in parameters).
\end{enumerate}

\begin{figure} \centering
    \includegraphics[width=1\linewidth]{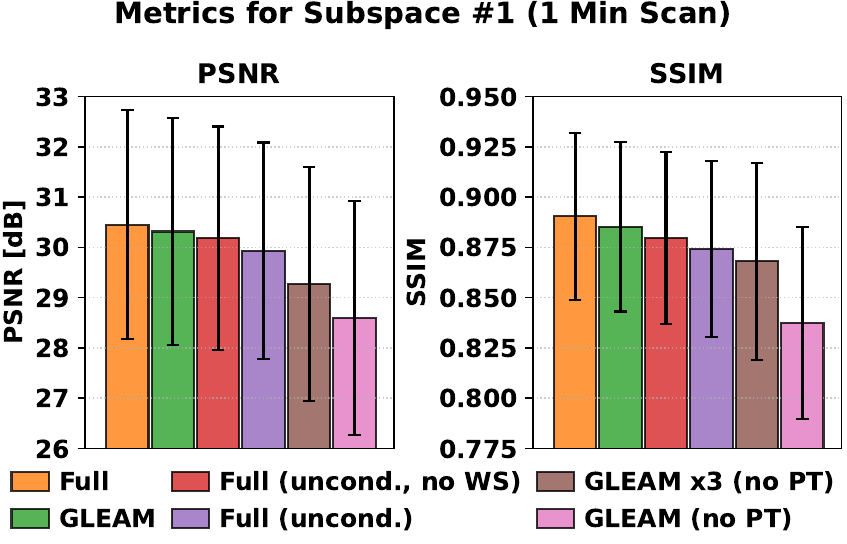}
    \caption{~Ablation of training components. PSNR (left) and SSIM (right) for the first subspace coefficient map in 1~min scans. Results are shown for the full approach (Full), training without final fine-tuning (GLEAM), GLEAM without pretraining (GLEAM no PT), GLEAM without pretraining but trained three times longer (GLEAM x3 no PT), full training without unroll iteration conditioning (Full uncond.), and full training without conditioning or weight sharing (Full uncond., no WS). Similar trends are observed across other scan durations and subspace coefficient indices.}
    \label{fig:training_ablation}
\end{figure}

Figure~\ref{fig:training_ablation} shows PSNR and SSIM for the different training
variants on 1~min scans (trends are similar for other durations). As can be seen, final fine-tuning provides additional improvements on top of GLEAM, while GLEAM training without pretraining performs poorly. Even when training GLEAM three times longer, performance does not reach that of GLEAM with pretraining.

\begin{figure} \centering
    \includegraphics[width=1\linewidth]{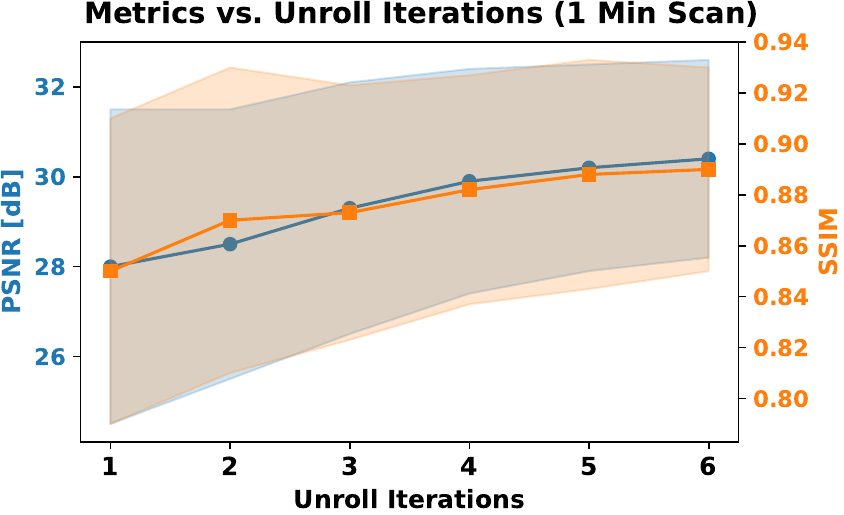}
    \caption{~PSNR and SSIM for the first subspace coefficient map in 1~min scans, for models trained with different numbers of unroll iterations. Similar trends are observed across other scan durations and subspace coefficients.}
    \label{fig:unroll_iters}
\end{figure}

Iteration conditioning improves performance, and when conditioning is removed, weight sharing decreases performance. Finally, Figure~\ref{fig:unroll_iters} reports PSNR and SSIM for models trained with a different number of unroll steps, showing that performance improves monotonically and converges around six unroll steps.

\subsection{Out-of-distribution generalization}
\begin{figure*} \centering
    \begin{subfigure}[t]{0.5\linewidth} \centering
        \includegraphics[width=\linewidth]{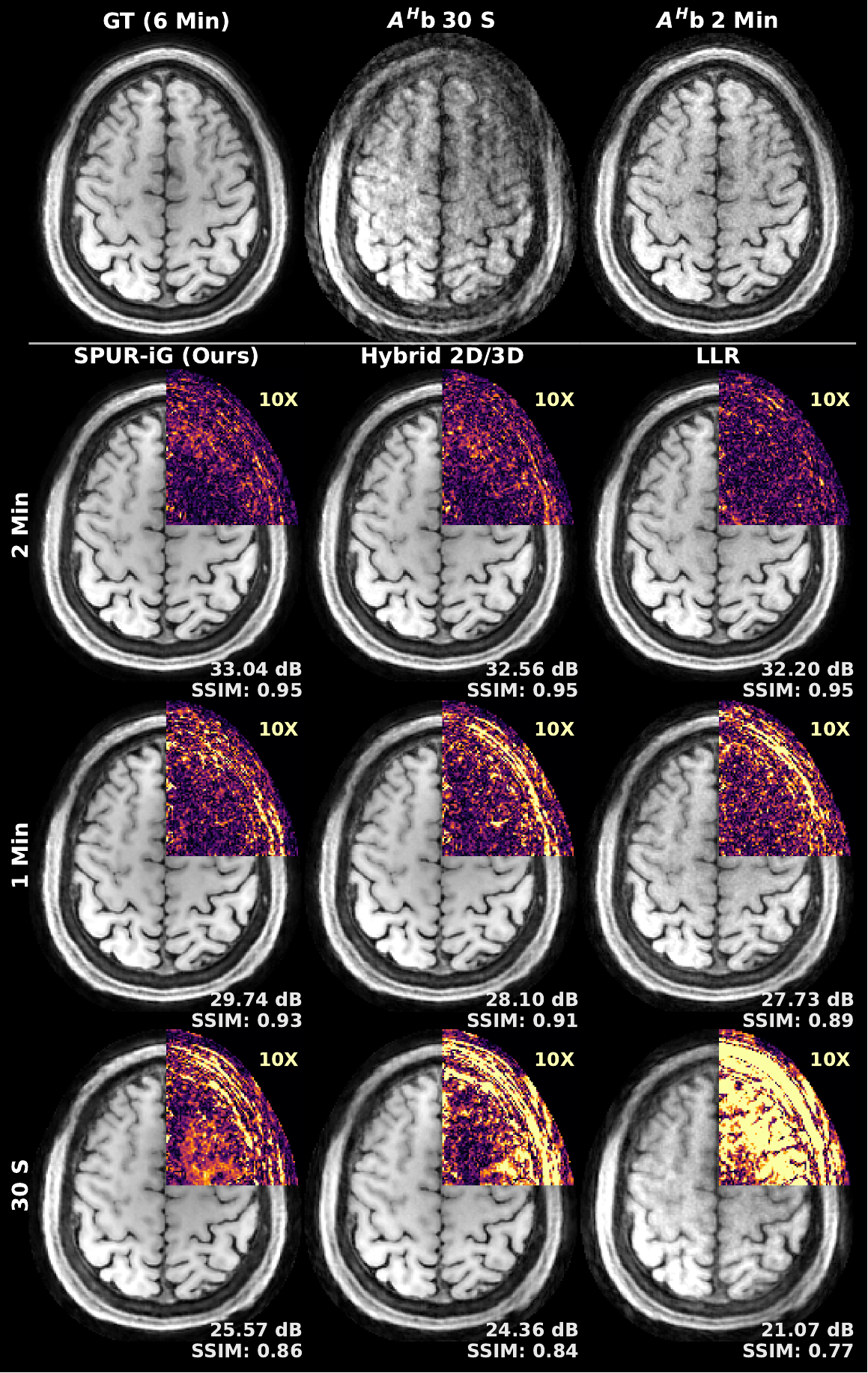}
        \caption{}\label{fig:siem_subspace_ex}
    \end{subfigure}
    \begin{subfigure}[t]{0.4915\linewidth} \centering
        \includegraphics[width=\linewidth]{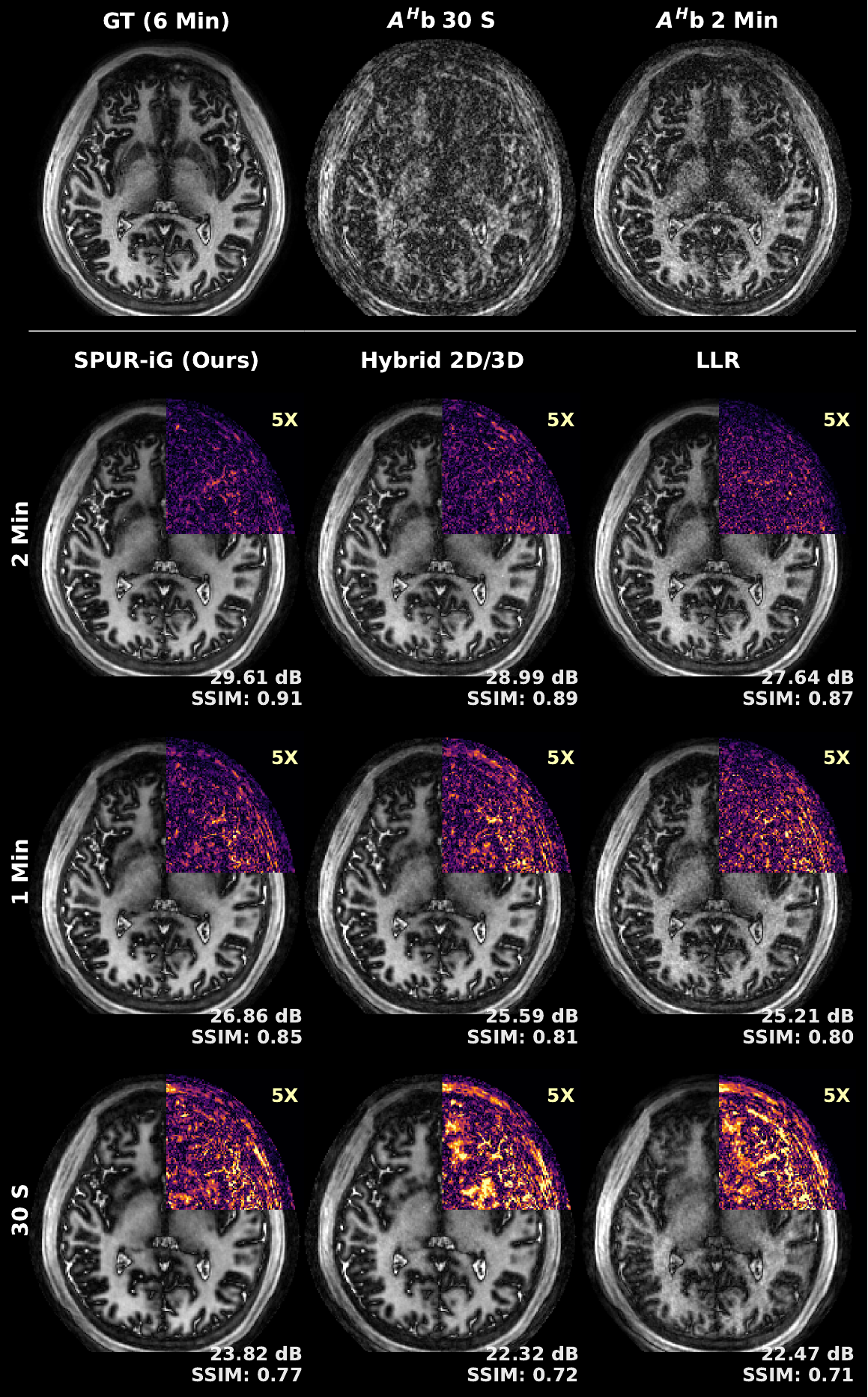}
        \caption{}\label{fig:siem_unbalanced_subspace_ex}
    \end{subfigure}
    \caption{~Reconstructions of the first balanced subspace coefficient (a) and the third unbalanced coefficient (b) for two slices from the same subject, scanned on an out-of-distribution scanner vendor. The layout follows Figure~\ref{fig:subspace_combined}.}\label{fig:siemens_subspace}
\end{figure*}

\begin{figure*} \centering
    \begin{subfigure}[t]{0.5\linewidth} \centering
        \includegraphics[width=\linewidth]{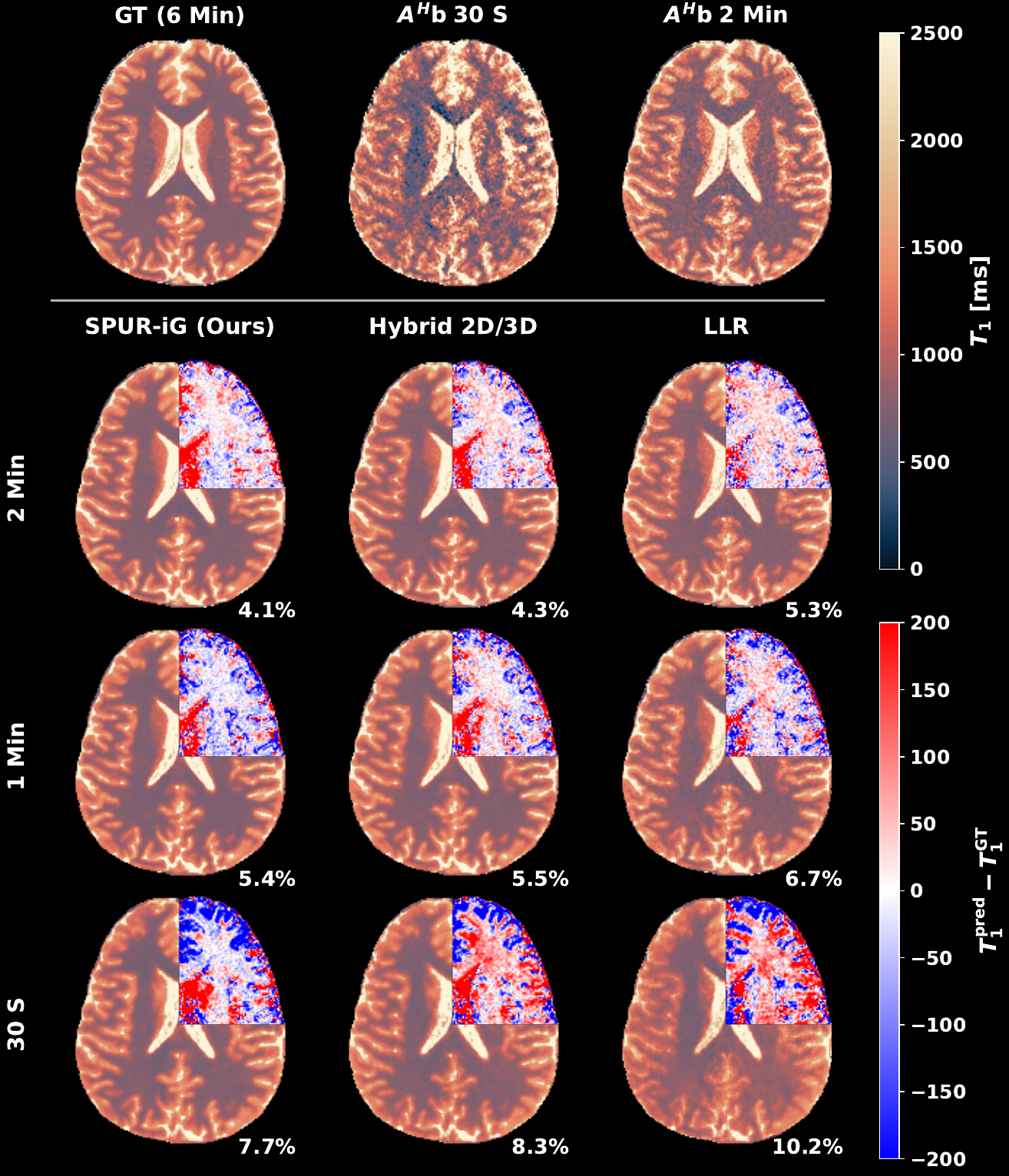}
        \caption{$T_1$ maps}
        \label{fig:quant_t1_ex_siem}
    \end{subfigure}
    \begin{subfigure}[t]{0.493\linewidth} \centering
        \includegraphics[width=\linewidth]{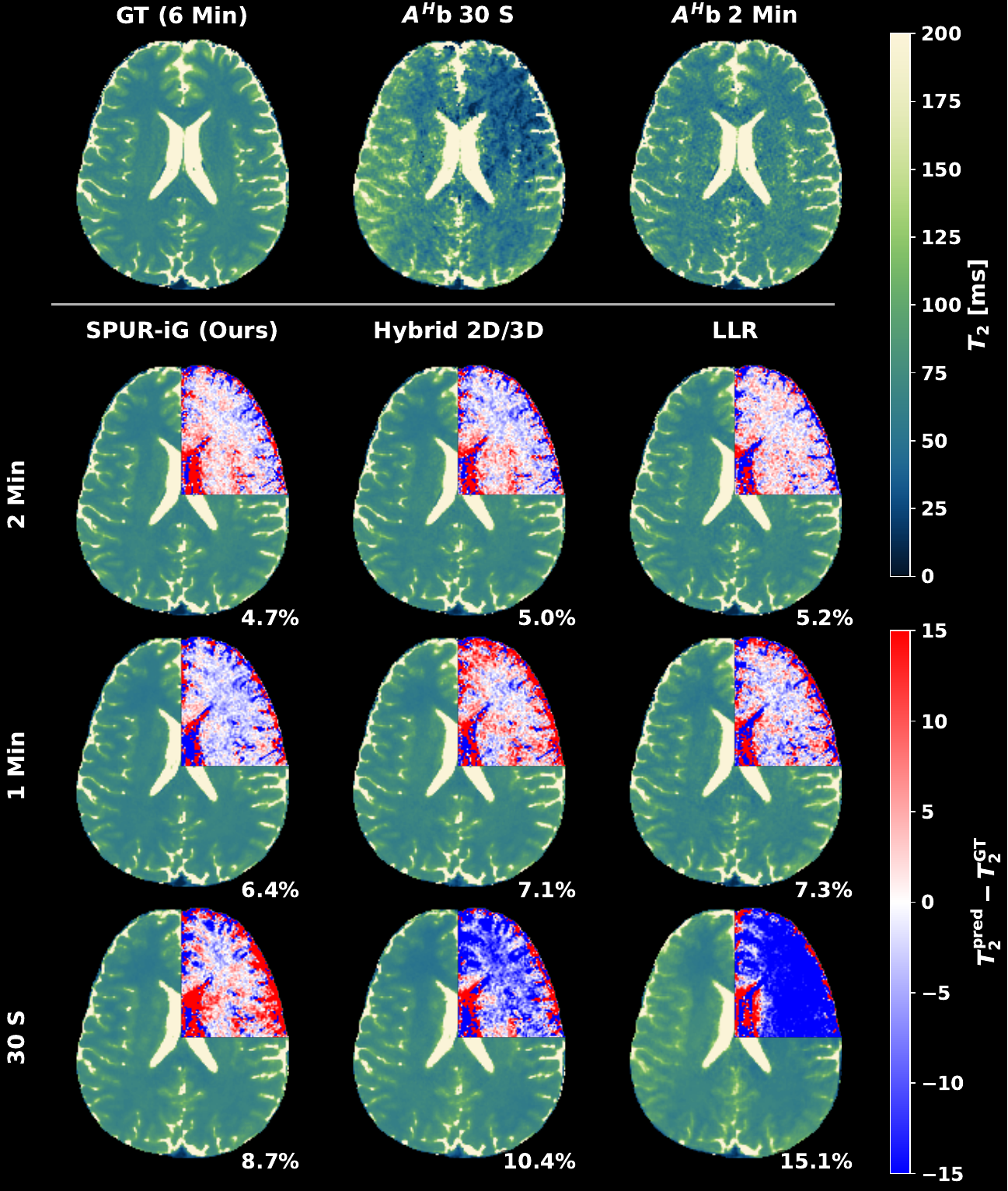}
        \caption{$T_2$ maps}
        \label{fig:quant_t2_ex_siem}
    \end{subfigure}
    \caption{~Qualitative examples of reconstructed $T_1$ (a) and $T_2$ (b) maps from an out-of-distribution vendor scan. The layout follows Figure~\ref{fig:quant_t1t2_ex}.}
    \label{fig:quant_t1t2_ex_siem}
\end{figure*}

To assess sensitivity to scanner vendor, we acquired data from a single subject on a Siemens 3T system using the same sequence and reconstructed it using the exact same model and setup (which was trained exclusively on GE data). Figures~\ref{fig:siemens_subspace} and~\ref{fig:quant_t1t2_ex_siem} show the corresponding subspace reconstructions and quantitative maps, which exhibit trends similar to those observed on the in-distribution test set.

\subsection{Reconstruction Time} 
\begin{figure} \centering
    \includegraphics[width=1\linewidth]{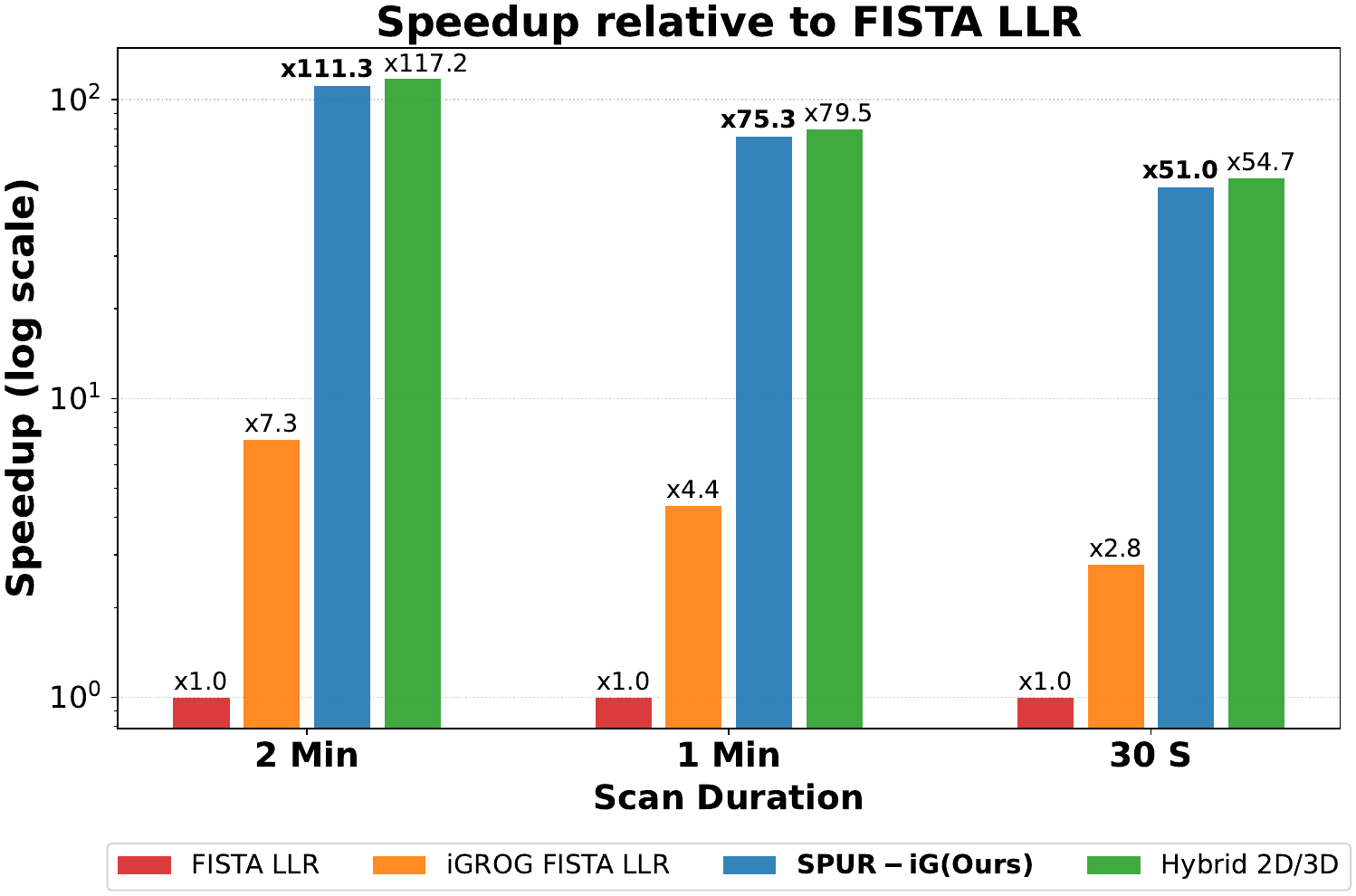}
    \caption{~Relative reconstruction speedup compared to FISTA LLR with 40 iterations. The chart compares iGROG with FISTA LLR (40 iterations), our method with six unroll steps, and the hybrid 2D/3D unrolled variant with six unroll steps. For methods using iGROG, reconstruction times include gridding but exclude kernel training (which takes less than 25~s, and can be performed in parallel to acquisition). For reference, the reconstruction times (in seconds) for each method and scan duration, from left to right, are: 1758.6, 241.7, 15.8, 15.0 (2~min); 986, 225.3, 13.1, 12.4 (1~min); and 601.7, 217.0, 11.8, 11.0 (30~s). All experiments are performed on a single A6000 GPU.}
    \label{fig:recon_times}
\end{figure}

Figure~\ref{fig:recon_times} summarizes reconstruction speedups relative to FISTA LLR. All experiments were performed on a Linux workstation using a single NVIDIA A6000 GPU (48GB memory) with a PyTorch implementation. Our method achieves up to $\times 111$ speedup for the 2~min acquisition and $\times 51$ for the 30~s acquisition. Compared to iGROG FISTA LLR, the speedups are $\times 15$ and $\times 18$ for the 2~min and 30~s cases, respectively. Notably, reconstruction completes in less than 15~s for all acquisition durations. We assume iGROG INR training is performed during MRF acquisition after calibration data are obtained and therefore does not add to reconstruction time, while gridding is included in the reported reconstruction times (kernel training completes in under 25~s and gridding in under 3~s across all acquisitions considered here). Speedup decreases for shorter scans as the DC cost scales with the size of the acquired k-space. The hybrid 2D/3D variant is marginally faster than the 3D model (by less than 1~s), but both provide comparable computational efficiency.

\section{Discussion}
\subsection{Quantitative Estimation}\label{sec:dis_quant}
With aggressive undersampling, LLR requires stronger regularization to suppress noise, which introduces systematic bias and produces overly smooth maps (Figure~\ref{fig:quant_t1t2_ex}, rightmost column). This is most evident at tissue boundaries, where the low-rank assumption does not hold. Although we jitter patch locations across iterations during LLR reconstruction to reduce blocking artifacts, the patch-based constraint still couples signals from different tissue types, leading to biased values at boundaries (e.g., elevated tissue values for WM and reduced values for GM or CSF). 

Our learned prior mitigates this effect by replacing patch-based low-rank constraints with a UNet denoiser. Unlike LLR, the UNet produces voxel-wise predictions informed by convolutional receptive fields, allowing it to incorporate spatial context while preserving local tissue differences. In practice, this enables the network to recognize and respect boundaries rather than enforcing a shared representation across them. Residual averaging effects remain, especially at shorter scan durations, but they are substantially smaller. The 3D model further reduces this bias by leveraging volumetric context. Most of the remaining error appears in CSF, which is expected since this particular MRF sequence is primarily designed to optimize sensitivity for WM/GM~\cite{cao2022optimized}.

Although our training objective is not explicitly designed to optimize tissue quantification, but rather to improve subspace coefficient reconstruction, Table~\ref{tab:quant_accuracy} shows that our method improves quantitative performance, highlighting the benefit of the learned prior. Future work could integrate quantitative parameter estimation directly into the pipeline and jointly optimize for estimation accuracy, or to optimize the learned prior for other downstream applications, such as multi-compartment modeling.

Finally, we use $B_1$ maps extracted from dummy group scans to correct $T_2$ bias when performing dictionary fitting. However, as scan duration decreases, the overhead from these groups becomes significant. Future work could integrate calibration-free methods to estimate $B_1$ maps directly from reconstructed coefficient maps~\cite{gao2023sequence}, reducing scan overhead and improving efficiency.

\subsection{Coefficient Maps}
Coefficient maps preserve richer temporal information than $T_1/T_2$ maps and are important for emerging downstream applications such as multi-compartment modeling and clinical contrast synthesis.  
As shown in Table~\ref{tab:combined_subspace}, our method consistently outperforms the baselines in PSNR and SSIM, with performance gaps widening as scan duration decreases. The improvements come from two factors: the learned prior, which captures more accurate image statistics than hand-crafted regularization, and the 3D UNet architecture, which leverages volumetric correlations across slices for stronger regularization. The proposed training strategy enables training and deployment of such a model, supporting a more complex and accurate representation of image statistics and, consequently, improved reconstructions.

The MRF scans used in this work utilize an optimized spatio-temporal k-space sampling strategy based on the TGAS approach~\cite{cao2022optimized}. Future work could further improve performance by refining sampling trajectories within the reconstruction training process~\cite{zou2025improved}, which may be particularly valuable for very short acquisitions such as 30~s.

\subsection{Method Design}
Figure~\ref{fig:training_ablation} evaluates the impact of different components of our framework:

\begin{enumerate}[leftmargin=*]
\item Pretraining is a key element of our approach. By avoiding unrolling and DC, it enables efficient training with extensive data augmentation, which improves generalization and robustness. Without pretraining, GLEAM performs poorly, and even tripling the training time is insufficient to reach comparable performance.

\item Fine-tuning with full unrolling provides a modest but consistent improvement by directly optimizing the final reconstruction loss. 
 
\item Conditioning on the unroll iteration index improves performance over unconditional variants. It simplifies the denoiser’s task by explicitly providing the expected artifact and noise-level distribution, rather than requiring the model to infer these solely from the input. In unconditional training, not sharing weights performs better than sharing, since the model no longer has to infer denoising strength from the input. However, both remain inferior to the conditioned version, which amortizes learning across noise regimes, effectively enlarging the dataset. The cost of conditioning is negligible, adding less than 1\% to the model parameters.
\end{enumerate}
These design choices enhance performance while achieving low runtime. The learned denoiser is effective at modeling complex distributions, and its inference is efficient, especially on modern GPUs, leading to reduced reconstruction times.  

Although the training set consists exclusively of data acquired on a 3T GE system, Figs.~\ref{fig:siemens_subspace} and~\ref{fig:quant_t1t2_ex_siem} demonstrate that the method generalizes to an out-of-distribution vendor without retraining or fine-tuning, indicating cross-vendor robustness. While some performance degradation is observed, we expect this gap could be mitigated by incorporating a small amount of out-of-distribution data during training or fine-tuning.

Finally, while demonstrated here for MRF, the proposed 3D unrolled training strategy is generally applicable to other large-scale imaging problems. Applications such as high-resolution structural and dynamic 3D imaging~\cite{ong2020extreme} may particularly benefit from its scalability and efficiency.

\section{Conclusions}
We propose a fully 3D unrolled MRF reconstruction framework with a staged training strategy that enables large-scale training within practical compute limits. By combining efficient iGROG-based DC with fast 3D UNet-based inference, the method achieves rapid reconstruction while consistently improving quality, supporting acquisitions as short as 30~s and preserving fidelity in both subspace coefficients and quantitative maps.

This represents an important step toward enabling short clinical acquisitions with fast reconstructions that can directly produce quantitative maps or synthesize clinical contrasts for routine use. Future work may explore alternative training objectives tailored to downstream applications, such as directly optimizing $T_1$/$T_2$ estimation or multi-compartment fitting. Another interesting direction is to extend the staged training strategy to even higher-resolution MRF such as $360\,\mu\text{m}$ isotropic \cite{cao2025invivo}, while keeping computation and memory demands tractable.

\vspace{-1em}\section*{Funding Information}\label{sec:funding_information}
\vspace{-1em}
Grant/Award Number: National Institutes of Health R01 EB033206, MH116173 \& EB019437.
\vspace{-0.5em}

\bibliography{reference}%

\clearpage
\vspace{-1em}\section*{Appendix}
\renewcommand{\thesubsection}{\Alph{subsection}}

\vspace{-0.75em}\subsection{Implementation Details}\label{sec:impl_details}\vspace{-0.5em}

\subsubsection{Model}
We implemented both 3D and 2D U-Nets with four encoder–decoder levels. Each level contained a residual block consisting of two convolutional layers with kernel size 3, followed by group normalization~\cite{wu2018group} (8 groups) and SiLU activations~\cite{elfwing2018sigmoid}. Additionally, a Squeeze-and-Excitation block~\cite{hu2018squeeze} with reduction factor 16 was applied after each residual block.

Inputs are $k=5$ complex subspace volumes, split into $C=10$ real/imaginary channels and whitened jointly for zero mean and unit covariance.

Iteration conditioning was implemented by mapping the unroll index $i$ to sinusoidal embeddings with dimension 48, followed by an MLP generating scale and shift parameters $(\gamma_i, \beta_i)$, injected into each block via FiLM-style modulation~\cite{perez2018film} $h' = (1 + \gamma_i) \odot h + \beta_i$.

Unrolled reconstruction is initialized with $\bcA^H \bb \cdot s$, where $s = \|\bb\|_2 / \|\bcA \bcA^H \bb\|_2$ is a scaling factor that aligns the norms of the initialization and the acquired k-space. We found this normalization reduces the number of required unrolled steps.

Training used Adam (initial learning rate $5 \times 10^{-3}$, halved after 20 epochs without validation improvement). Pretraining required 4~GPU days, GLEAM training 2~GPU days, and fine-tuning 2~GPU days, all on a single NVIDIA A6000 (48~GB).  

\subsubsection{Pretraining}
In pretraining, input types were heuristically matched with iteration indices as
\begin{align*}
\bcA^H \bb &\;\mapsto\; 1, \\
\balpha_{\text{LLR}}^{\text{30s}} &\;\mapsto\; 2, \\
\balpha_{\text{LLR}}^{\text{1min}} &\;\mapsto\; 2, \\
\balpha_{\text{LLR}}^{\text{2min}} &\;\mapsto\; 3, \\
\balpha_{\text{LLR}}^{\text{6min}} &\;\mapsto\; k,\quad k \sim \mathcal{U}[4,6].
\end{align*}
Here, the highest-quality input $\balpha_{\text{LLR}}^{\text{6min}}$, which also serves as the ground truth, is assigned a randomly sampled iteration index $k$ drawn uniformly from $\mathcal{U}[4,6]$. Data augmentations included random rotations ($\leq 15^\circ$), translations (up to $\pm5$ voxels), and scaling in $[0.98,1.02]$. To increase batch diversity, pretraining used randomly cropped $96^3$ patches with a batch size of 8. 

\paragraph{Tissue Aware Loss}\label{sec:apx_loss}
We define a tissue-aware spatially weighted $L_1$ which we use during pretraining:
\begin{equation}
\mathcal{L}_{\text{spatial}}(\hat{\balpha}, \balpha)
= \sum_{v \in \{\text{WM}, \text{GM}, \text{CSF}, \text{Other}\}} \lambda_v \, \lVert \hat{\balpha}_v - \balpha_v \rVert_1,
\end{equation}
where $\balpha_v$ and $\hat{\balpha}_v$ denote the ground truth and estimated coefficient images restricted to tissue class $v$ (obtained from the segmentation masks computed from the GT reconstruction), and `Other` corresponds to voxels outside the WM/GM/CSF masks. We set $\lambda_{\text{WM}} = \lambda_{\text{GM}} = 10$ and $\lambda_{\text{CSF}} = \lambda_{\text{Other}} = 1$. This weighting reflects the fact that the acquisition protocol is specifically optimized for sensitivity to WM and GM, and we therefore prioritize accurate estimation in these tissue classes.

\subsubsection{GLEAM stage}
In the GLEAM stage, supervision is applied to all intermediate unroll outputs $\{ \balpha_k \}_{k=1}^K$ using geometrically increasing weights, such that the highest weight is assigned to the final output at step $K$. Additionally, we augment the $L_1$ loss with a multi-scale structural similarity (MS-SSIM). The loss is defined as
\begin{align}
\mathcal{L}_{\text{GLEAM}}
&= \sum_{k=1}^{K} \lambda_k \, \mathcal{L}_{\text{spatial}}(\balpha_k, \balpha_{\text{gt}}) \\
&\quad + \lambda_{\text{SSIM}} \, \mathcal{L}_{\text{MS-SSIM}}(\balpha_K, \balpha_{\text{gt}}),
\end{align}
where $\balpha_{\text{gt}}$ denotes the ground truth, and $\{\lambda_k\}$ follow a geometric progression with $\lambda_K = 10 \lambda_1$. This weighting scheme encourages progressive refinement across unroll iterations while emphasizing accuracy of the final reconstruction. Due to memory constraints, MS-SSIM is computed on 30 randomly selected axial slices per volume. Here we use full volumes as input with a batch size of 1, hence no stitching logic is required, and employed gradient accumulation with a factor of 4.

\subsubsection{Fine-tuning}
The fine-tuning stage uses the same loss, batch size and gradient accumulation formulation as GLEAM, but supervision is applied only to the final output.

\subsubsection{LLR Reconstruction}
Reconstruction used $10^3$ patches with random shifts per iteration, and SVD thresholds of $5 \times 10^{-5}$ (6~min), $1 \times 10^{-4}$ (2~min, 1~min), and $1.5 \times 10^{-4}$ (30~s).

\end{document}